\documentclass[
amsmath, 
amssymb, 
aps, 
pra,
floatfix,
nofootinbib
]{revtex4}

\usepackage[utf8]{inputenc}
\usepackage{titlesec}
\usepackage[colorlinks]{hyperref}
\usepackage{amsmath,amssymb}
\usepackage{bbold}
\usepackage{float}
\usepackage{bm}
\usepackage{url}
\usepackage{bbold}
\usepackage{makecell}
\usepackage[normalem]{ulem}
\usepackage{xcolor}

\usepackage{csquotes}

\begin{document}
\title{Applicability of QKD: Terra Quantum view on the NSA’s scepticism}
\author{D.\,Sych\footnote{These authors were previously affiliated with Terra Quantum and currently are independent researchers}}
\author{A.\,Kodukhov}
\author{V.\,Pastushenko}
\author{N.\,Kirsanov}
\author{D.\,Kronberg$^*$}
\author{M.\,Pflitsch}
\address{Terra Quantum AG, Kornhausstrasse 25, 9000 St. Gallen, Switzerland
\\
Correspondense should be sent to ak@terraquantum.swiss (A.K.)}

\begin{abstract}
Quantum communication offers unique features that have no classical analog, in particular, it enables provably secure quantum key distribution (QKD). Despite the benefits of quantum communication are well understood within the scientific community, the practical implementations sometimes meet with scepticism or even resistance. In a recent publication \,\cite{NSA}, NSA claims that QKD is inferior to ``quantum-resistant'' cryptography and does not recommend it for use. Here we show that such a sceptical approach to evaluation of quantum security is not well justified. We hope that our arguments will be helpful to clarify the issue.
\end{abstract}
     
\maketitle

\bigskip
\begin{displayquote}
\textit{\textbf{NSA claims: \\Quantum key distribution is only a partial solution}}

\textit{QKD generates keying material for an encryption algorithm that provides confidentiality. Such keying material could also be used in symmetric key cryptographic algorithms to provide integrity and authentication if one has the cryptographic assurance that the original QKD transmission comes from the desired entity (i.e. entity source authentication). QKD does not provide a means to authenticate the QKD transmission source. Therefore, source authentication requires the use of asymmetric cryptography or preplaced keys to provide that authentication.
}\end{displayquote}

Authentication is outside of the QKD scope, since there is no way to prevent a man-in-the-middle attack, neither quantum nor classical, without authentication. Indeed, in order to assure the identity of two remote parties, they have to know something unique about each other, i.e. pre-distribute secure keys, either directly or indirectly via a trusted third party.
Authentication procedure itself can be realised in an information-theoretically secure way (e.g. 2-universal hash function\,\cite{krovetz_hash, EIST}), which does not compromise the rigorously provable QKD security statement, and complements the further use of quantum keys in informationally-secure classical cryptography.

Classical encryptors also require prior trusted key distribution and periodic key changes in accordance with NIST standards for NSS. In comparison to classical cryptography, QKD offers much simpler and effective distribution of these keys: the initial long-term service keys are distributed together with the QKD transceiver devices, and the long-term and session keys for consumers are generated along with the quantum keys. This allows to change the key without moving the key carrier and reduces the human factor. It should be noted that the more important the information processed in the NSS, the more often the encryption key should be changed. Moreover, the everlasting security feature of QKD makes any attempts to collect information about the keys and use it in the future irrelevant\cite{renner2023everlasting}.

\begin{displayquote}
\textit{\textbf{NSA claims:}}

\textit{
Moreover, the confidentiality services QKD offers can be provided by quantum-resistant cryptography, which is typically less expensive with a better understood risk profile.
}\end{displayquote}

The statement is misleading.
A durable key is typically used to distribute or co-produce a session key.
This can happen with symmetric and asymmetric keys, but regardless of key types, each key has a lifetime at the end of which it is changed.
When network nodes are geographically distributed such key replacement can be quite expensive as it may require physical delivery of a new key.
The use of QKD can alleviate this problem.
Security of QKD is based on fundamental laws of quantum physics and rigorous mathematical proofs \cite{Shor_preskill_security}.
In contrast, classical cryptography has none of these features and relies on unproved computational complexity. Despite the fancy name, there is no guarantee that ``quantum-resistant cryptography'' won’t be broken in the near future (if not already) by a classical computer algorithm via a clever design of latter (or former).
Thus a quantum key can be combined with a classical key, which reduces the load and risks on each of them. 

\bigskip
 \begin{displayquote}
\textit{\textbf{NSA claims:\\
Quantum key distribution requires special purpose equipment}}

\textit{
QKD is based on physical properties, and its security derives from unique physical layer communications. This requires users to lease dedicated fiber connections or physically manage free-space transmitters. It cannot be implemented in software or as a service on a network, and cannot be easily integrated into existing network equipment. Since QKD is hardware-based it also lacks flexibility for upgrades or security patches.
}\end{displayquote}

Indeed, QKD requires special purpose equipment and cannot be implemented as a software service only. QKD is not meant to be a universal replacement for classical security, but is mostly targeted to protecting the most sensitive data. Albeit the classical cryptography can be realised as a special software running on conventional computers, it worth to mention that the computers are highly sensitive to the numerous hacking attacks, and the practical encryption of the sensitive data is performed on the dedicated equipment. Taking into account the overall price of the complete system, the quantum hardware does not add significant costs. Moreover, the price tag of modern QKD systems comes due to the limited production rather than due to objectively expensive components, and will certainly drop down as market will grow.

Integration of QKD into the existing network infrastructure is not easy, though possible. The main obstacle is fragility of quantum optical signals (which is the basis for the QKD security). Several groups have already demonstrated the coexistence of classical and quantum communication over the same optical fibre via WDM \cite{wdm1, wdm2}, although the use of a dedicated fibre just for QKD is not something overwhelmingly difficult or obstructive.

Finally, the statement regarding the lack of flexibility for upgrades is incorrect. In fact, the software security patches that come regularly on a variety of electronic devices have to be protected themselves. If a QKD equipment requires such patches, the most safe way to deliver them in full integrity is to use the quantum keys generated by the QKD equipment itself. 

\bigskip
 \begin{displayquote}
\textit{\textbf{NSA claims:\\
Quantum key distribution increases infrastructure costs and insider threat risks}}

\textit{
QKD networks frequently necessitate the use of trusted relays, entailing additional cost for secure facilities and additional security risk from insider threats. This eliminates many use cases from consideration.
}\end{displayquote}

The statement is misleading. The necessity of trusted nodes, where quantum information is converted into the classical form and can be potentially replicated, concerns only the long-distance networks. The modern world-record distances of QKD communication over the optical fibres without the trusted nodes is of the order of 1000 km\,\cite{TF}, which is clearly sufficient for many use cases. In particular, this is clearly sufficient to protect channels between geographically distributed data centers, backbone provider lines, corporate networks, intracity links, to name a few.

To enlarge the span of the secure links beyond 1000 km, quantum communication offers several solutions. First, development of novel quantum communication protocols and quantum protection procedures enables the Earth-scale links even over the optical fibres \cite{forty, boosting, aliev2023experimental}. Second, development of quantum memory enables quantum repeaters that expand the secure communication distance in an information-theoretically secure way without any limit on the maximal distance. Last but not least, the use of satellite free-space links instead of optical fibres greatly expands the communication distance as well \cite{free_space1, free_space2}.

\bigskip
 \begin{displayquote}
\textit{\textbf{NSA claims:\\
Securing and validating quantum key distribution is a significant challenge}}

\textit{
The actual security provided by a QKD system is not the theoretical unconditional security from the laws of physics (as modeled and often suggested), but rather the more limited security that can be achieved by hardware and engineering designs. 
}\end{displayquote}

To be correct, we agree with the more refined statement: The actual security provided by any cryptographic system (either classical, or quantum, or post-quantum) is not the theoretical security, but rather the more limited security that can be achieved by hardware and engineering designs. The problem is much more complicated than it looks from the first sight. Security is an object from the real world, while a cryptographic protocol, to which information security proofs can be applied, is an abject from the idealised world of theoretical models. These two worlds deviate (and always will) from each other, which results in loopholes and hacking \cite{Makarov}. From the practical point of view, if a given cryptographic system is hacked, there is no difference for the users whether the hacking is due to the loopholes in the theoretical model or due to the loopholes in hardware. Classical cryptographic systems do have both kind of loopholes, while quantum cryptographic systems are proven to have no theory loopholes, at least. Moreover, the quantum theory provides the ways to close a large class of hardware loopholes by the proper design of a communication protocol. E.g. MDI closes a large class of vulnerabilities within the receiver side \cite{mdi1, mdi2, mdi3}, and DI closes both receiver and transmitter hardware loopholes \cite{DI1,DI2}. These features have no classical analog and make quantum communication much more resistant to the hardware imperfections. At least the quantum theory provides the ways to do this. 

\begin{displayquote}
\textit{\textbf{NSA claims:}}

\textit{
The tolerance for error in cryptographic security, however, is many orders of magnitude smaller than in most physical engineering scenarios making it very difficult to validate. The specific hardware used to perform QKD can introduce vulnerabilities, resulting in several well-publicized attacks on commercial QKD systems.
}\end{displayquote}

The statement is misleading. Errors in quantum and classical communication devices are of different kinds. Usually classical communication implies digital communication with the well-defined signal levels, while quantum communication deals with the well-defined quantum properties of optical signals, such as quantum uncertainty or quantum entanglement. Both types do have imperfections, and, consequently, the impact on security, but their direct comparison is impossible. What is possible though, is, for example, to consider an error in voltage of, say, a signal generator, and analyse how it affects the security of a system in the worst case scenario. Indeed, such technical vulnerabilities affect the practical security of QKD, but they also affect the security of classical cryptographic systems as well. Moreover, the QKD security deals with the $\varepsilon$-security framework \cite{epsilon}, which includes all kinds of possible errors, and may also include estimates of hardware failures and guessing the whole key by the eavesdropper. Limited reliability of some equipment still makes it possible to estimate the overall system security, and upgrade the weak parts of the system.

It worth to mention here, that the ever-growing list of attacks on classical electronic devices is orders of magnitude larger compared to the list of attacks on QKD. E.g. the well-known processor vulnerabilities like Spectre or Meltdown allow to read the secret information, like cryptographic keys, albeit the later are distributed by a whatever secure system\,\cite{Spectre, Meltdown}.

\bigskip
\begin{displayquote}
 \textit{\textbf{NSA claims:\\
 Quantum key distribution increases the risk of denial of service}}

\textit{
The sensitivity to an eavesdropper as the theoretical basis for QKD security claims also shows that denial of service is a significant risk for QKD.
}\end{displayquote}

Indeed, the current point-to-point realisations of QKD are sensitive to the noise in communication channel. There is no problem for a sufficiently smart eavesdropper to inject a noisy optical signal in the optical fibre and ruin the QKD session (but not to obtain the secret key!). Such kind of DDoS attacks can be overcome by the development and implementation of noise-resistance QKD protocols \cite{noise}, as well as by the development of quantum networks, where the key is routed though many physical fibres \cite{network1, network2, network3}. 

In addition, DDoS attacks are not something unique to quantum communication. Actually, a denial of service of classical communication systems is not very rare, as one can see from the daily practice.

\bigskip
\begin{displayquote}
\textit{\textbf{NSA claims:\\
Conclusion}}

\textit{
In summary, NSA views quantum-resistant (or post-quantum) cryptography as a more cost effective and easily maintained solution than quantum key distribution. For all of these reasons, NSA does not support the usage of QKD or QC to protect communications in National Security Systems, and does not anticipate certifying or approving any QKD or QC security products for usage by NSS customers unless these limitations are overcome.
}\end{displayquote}

We show that most of the NSA’s concerns about the technical limitations of QKD (e.g. the need for pre-distributed keys, pricy special purpose equipment, flaws and side channels in hardware and engineering, denial of service, etc) are also applicable to the classical (in particular, to post-quantum) information security systems. To avoid any misjudgement and make a fair comparison of classical and quantum security systems, this should be clearly stated. The other concerns (e.g. the limited communication range, a less developed risk profile, additional costs, etc) are reasonable, but do not place severe restrictions on the use of QKD. The field of quantum communication is relatively young compared to the classical information security, especially from the perspective of practical issues and technical imperfections. There is no doubt that such problems are solvable within the quantum approach. Here we note that not all standardisation services share the NSA's skepticism. For example, ETSI, ITU-I and other standards for QKD have already been adopted.

Finally, we would like to point out that the development of quantum communication provides the unique instruments to guarantee maximum and everlasting security for the most sensitive information — the feature with no classical analog\,\cite{renner2023everlasting}. Thus post-quantum cryptography cannot replace QKD but rather complements it in the real-world applications, e.g. the long-term key and the authentication key are quantum, while the session key and encryption can be classical. We hope that the arguments provided in this manuscript, as well as in the previous manuscripts on this issue\,\cite{renner_debate}, pave the way to overcome the NSA’s concerns about the quantum communication.

\bibliography{re.bib}

\end{document}